\documentclass{siamart171218}
\usepackage{comment}
\usepackage{amsfonts}
\usepackage{graphicx}
\usepackage{epstopdf}
\usepackage{algorithmic}
\ifpdf
  \DeclareGraphicsExtensions{.eps,.pdf,.png,.jpg}
\else
  \DeclareGraphicsExtensions{.eps}
\fi

\title{A Modified Bisecting K-Means for Approximating Transfer Operators: Application to the Lorenz Equations\thanks{Submitted to the editors DATE.
\funding{This work acknowledges support by Schmidt Sciences, LLC, through the Bringing Computation to the Climate Challenge, an MIT Climate Grand Challenge Project.}}}

\author{Andre N. Souza\thanks{EAPS, MIT, Cambridge, MA 
  (\email{andrenogueirasouza@gmail.com}, \url{sandreza@github.io})}
  \and Simone Silvestri\thanks{EAPS, MIT, Cambridge, MA 
  (\email{ssilvest@mit.edu}).}
}

\ifpdf
\hypersetup{
  pdftitle={A Modified Bisecting K-Means for Approximating Transfer Operators: Application to the Lorenz Equations},
  pdfauthor={Andre N. Souza and Simone Silvestri}
}
\fi

\begin{document}

\maketitle

\begin{abstract}
We investigate the convergence behavior of the extended dynamic mode decomposition for constructing a discretization of the continuity equation associated with the Lorenz equations using a nonlinear dictionary of over 1,000,000 terms. The primary objective is to analyze the resulting operator by varying the number of terms in the dictionary and the timescale. We examine what happens when the number of terms of the nonlinear dictionary is varied with respect to its ability to represent the invariant measure, Koopman eigenfunctions, and temporal autocorrelations. The dictionary comprises piecewise constant functions through a modified bisecting k-means algorithm and can efficiently scale to higher-dimensional systems. 
\end{abstract}

\begin{keywords}
  Fokker-Planck Equation, Continuity Equation, Lorenz Equations
\end{keywords}

\begin{AMS}
  68Q25, 68R10, 68U05
\end{AMS}

\section{Introduction}

In the absence of deterministic predictability, we are often relegated to making statistical statements about the system in question. The lack of predictability can arise due to the intrinsic stochasticity of an underlying equation, the lack of information on precise initial conditions, uncertainty in model parameters, or the non-existence of fundamental guiding physical principles to formulate an evolution equation. In addition to the necessity of a probabilistic perspective in the former cases, switching to a probabilistic perspective offers insight into known deterministic equations with precisely known initial conditions, especially in chaotic and ergodic dynamics. 

Our work focuses on applying a modified bisecting k-means algorithm to test the limits of the extended dynamic mode decomposition in representing spectral properties of a chaotic dynamical system. We test our methodology on the Lorenz equations \cite{lorenz1963deterministic} and pose several conceptual questions: 
\begin{enumerate}
    \item What does it mean to approximate the Koopman eigenfunctions of a chaotic dynamical system in the absence of noise?
    \item To what extent does increasing the number of terms of a nonlinear dictionary ``help"? 
    \item At what timescale(s) should a Koopman operator be constructed? 
\end{enumerate}
Concretely, there is an associated Liouville / Fokker-Planck equation to every dynamical system that describes the evolution of a probability density in state space. We construct a data-driven discretization of this equation by leveraging the time series of its dynamical evolution through an extended dynamic mode decomposition. We use a nonlinear dictionary of piecewise constant functions with over 1,000,000 terms to construct a discretization of the Fokker-Planck operator (or, equivalently, the Koopman operator). 

Instead of focusing on the dynamics of an evolution equation, we focus on the statistics. When the evolution equation is given by a set of $n$ ordinary differential equations, this amounts to an evolution of an $n$-dimensional partial differential equation. As such, there are many standard methods that can be employed for discretizing a partial differential equation in a low number ($\leq 4$) of dimensions; however, we wish to employ an extended dynamic mode decomposition to extract a discretization from the dynamics of a time series such as \cite{Ulam1964, Froyland2001, Dellnitz2005}. Traditional methods are applicable for the stochastic Lorenz equations, \cite{Marston2016}, but modern methods \cite{Schmid2010, koopman_2009, Klus2016, GIANNAKIS2019338, Fernex2021} with rigorous convergence guarantees \cite{Froyland1997, DAS202175, ColbrookTownsend2021, SchuetteKlusHartmann2022} are more akin to what will be performed here. Deep learning methods to learn optimal nonlinear dictionaries for observables are also beginning to make progress, see \cite{constanteamores2023enhancing}  and \cite{lump_2023}.

What we present here is an efficient method for constructing Koopman operators,  \cite{MezicReview2013, Froyland2013, Williams2015, Rosenfeld_2021, Fernex2021}. As noted in \cite{Souza_2024_1, Souza_2024_2}, using a piecewise constant basis for the nonlinear dictionary allows for computational expedience, which we further accelerate by applying a modified bisecting k-means algorithm. The only limitation of the algorithm is the availability of sufficient data. Other efficient methods have been constructed in the past, see \cite{Dellnitz_1999, Dellnitz_2001, Koltai_2009}, but have been limited to structured assumptions that are not present in this work. All computations performed in the present work was performed a single cpu core. 

Our methodology can be applied to any dynamical system; see \cite{Souza_2024_1, Souza_2024_2} for an example involving the Lorenz equations and the compressible Euler equations on the sphere, but we focus on the Lorenz equations to directly address challenges associated with the representation of Koopman operators in a simplified setting. Preliminary versions of the algorithm here was implemented already for higher-dimensional dynamical (128+ dimensional) systems in \cite{giorgini2023clustering}.  Past studies of the Lorenz equation in the statistical setting include \cite{allawala2016statistics} where a direct discretization of the Fokker-Planck operator of the stochastic Lorenz equations was implemented. 

We modify the ``bisecting k-means" clustering algorithm, see \cite{bisecting_kmeans}, and apply it to  constructing a discretization the continuity equation associated with the Lorenz equations. The k-means algorithm, initially introduced in \cite{macqueen1967methods}, has been extended and adapted in various ways, including dynamic modeling for hierarchical clustering \cite{karypis1999chameleon, JAIN2010651}. The primary difference with our method is that a splitting criterion is introduced, and a cluster is only split if an additional auxiliary condition is met. The result of our chosen criteria is that the partitions have nearly uniform entropy. Thus, unlike k-means, the number of clusters is not specified apriori but rather a fixed probability, $p$, which empirically yields $k \approx 1.5/ p$ clusters for the calculations here. 

At the end of applying the method, one has an appropriate ``classifier" for a given dynamical system state, from whence the construction of the Perron-Frobenius Operator (or its generator) follows, see \cite{Souza_2024_1}. We investigate the effect of data sampling frequency and varying the number of clusters (cells) in a partition. We assess the operator's convergence in terms of its ability to represent the invariant measure, Koopman eigenfunctions, and the spectra, drawing on the spectral properties of dynamical systems as discussed by
 \cite{mezic2005spectral}.

This paper is organized as follows: Section \ref{sec:algorithm} describes the algorithm, Section \ref{sec:lorenz} covers the application of the algorithm to the Lorenz equations, and Section \ref{sec:conclusions} concludes with future directions.

\section{Algorithm}
\label{sec:algorithm}

We first review the bisecting k-means algorithm as succinctly stated in \cite{bisecting_kmeans}: 
\begin{enumerate}
    \item Choose a cluster to split 
    \item Find two sub-clusters using the basic k-means algorithm (bisect)
    \item Repeat step 2 for $N$ times and take the split that produces the clustering with the highest overall similarity
\item Repeat 1, 2, 3 until the desired number of clusters is reached.
\end{enumerate}
We make two modifications to this algorithm in order to construct a ``classifier" for new data 
\begin{enumerate}
    \item Introduce a splitting criteria to stop applying k-means to a cell. 
    \item Apply the algorithm to a subset of the full dataset. 
\end{enumerate}
Thus, the modified algorithm is 
\begin{enumerate}
 \item Determine whether or not a cluster should be split according to a chosen criteria 
\item Find sub-clusters using the basic k-means algorithm 
\item Repeat steps 1 and 2 until there are no more clusters to split 
\end{enumerate}
All computations in the present work use Euclidean distance as the notion of similarity (where a smaller distance between points is meant to represent states that are more ``similar") since the operations are performed in the low-dimensional setting. 

Unlike k-means, we do not specify the number of clusters, $k$,  a-priori, but instead specify a maximum number of points that can exist within a cluster. If the number of points within a cluster exceeds this threshold, we apply k-means again within a given cluster. Of course, one could specify any criteria in order to determine whether or not to further split the cluster, but choosing a ``maximum number of points within the cluster" criteria quickly yields a nearly uniform entropy at the cost of an unbalanced tree, a sacrifice that we are willing to make. This choice can now be tied to our second modification. In practice we specify ``minimum probability" for a cluster, $p_{\text{min}}$. The number of data points necessary for all the clusters to be below this threshold is estimated as $N_{\text{min}} = \text{floor}(100 / p_{\text{min}})$.  If $N_{\text{min}}$ is larger than the number of data points $N_{\text{data}}$ we instead set $p_{\text{min}} = 100 / N_{\text{data}}$. The ``maximum points in a cluster" criteria is then set to $N_{\text{threshold}} = \text{ceil}(p_{\text{min}}  N_{\text{min}})$. Thus, when k-means is applied, if the number of points in a cluster $N_{\text{cluster}}$ is larger than $N_{\text{threshold}}$, a splitting criteria is applied. In principle, one could determine how to split the data with only 100 or so points within a cluster, but instead, we start with a sufficiently large subset of the original data to apply the splitting. 

\begin{figure}
\begin{center}
\includegraphics[width=1.0\textwidth]{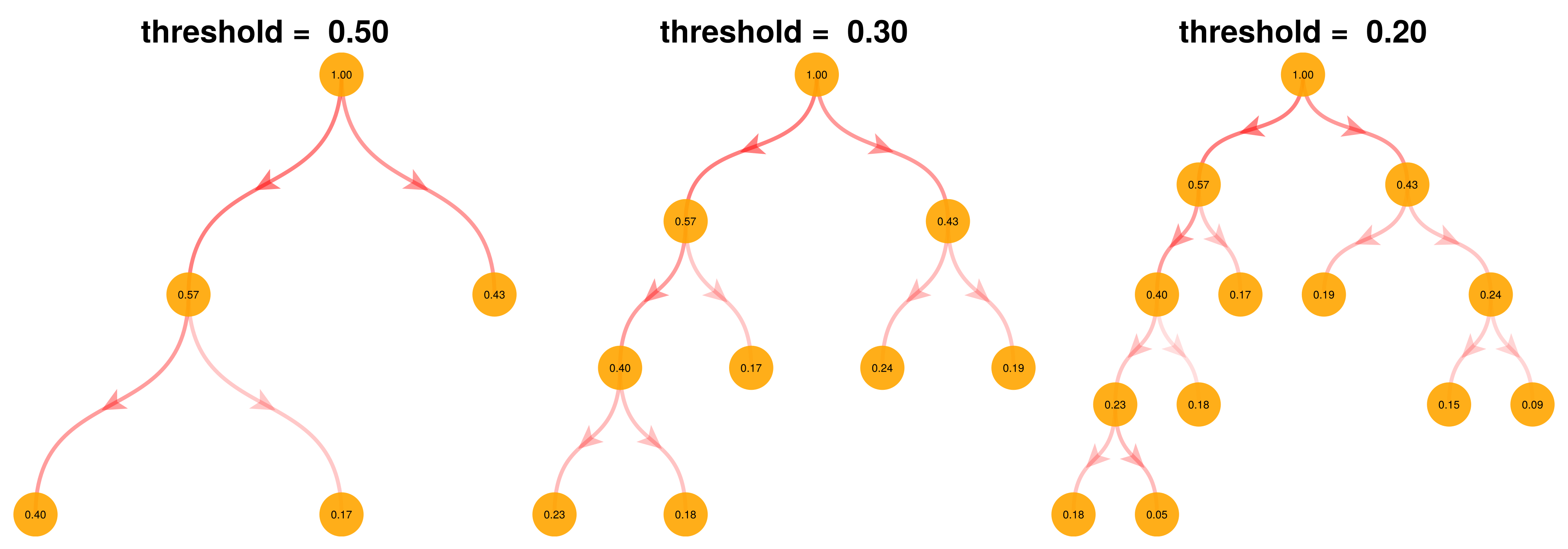}
\caption{\textbf{Modified Bisecting K-Means Algorithm for Different Thresholds.} The number on a node represents the probability of finding a data sample within the cluster. At each node, the threshold criteria is checked to determine whether or not a bisection should be performed. The leaf nodes are the resulting clusters. }
\label{fig:tree}
\end{center}
\end{figure}

The procedure is illustrated in Figure \ref{fig:tree}. On the left tree, we start with a probability threshold of $p = 0.5$. After the first split, the two nodes contain $57 \%$ and $43 \%$ of the data. The $43 \%$ node falls below the $50 \%$ threshold; thus, no further splitting is necessary. On the other hand, $57 \% \geq 50 \%$ and another level of splitting occurs. The resulting center nodes divide the $57\%$ probability into two additional clusters that hold $40 \%$ and $17 \%$ of the data, respectively. Since both resultant nodes fall below the $50 \%$ threshold, splitting stops, and we are left with three clusters that contain $40 \%$, $17 \%$, and $43 \%$ of the data. Approximately $43\%$ of the time, we only need to evaluate two similarity (distance in the present case) functions, and $57 \%$ of the time, we need to evaluate four similarity functions, leading to an average of $3.14 \approx 2  \times 0.43 + 4 \times 0.57$ similarity function evaluations on average. This value is suboptimal for such a low number of clusters, but the efficiency reveals itself when considering the $20 \%$ threshold on the right. There are seven leaf nodes, and the worst-case scenario is eight similarity evaluations which happen $23 \%$ of the time. In total, the average number of similarity evaluations is $5.8 \approx 8 \times 0.23 + 4 \times 0.17 + 6 \times 0.18 + 4 \times 0.19 + 6 \times 0.24$, which already shows a benefit over naively evaluating a similarity across seven different categories. 

After one has obtained a classifier with the method, one must perform additional ``similarity evaluations" for new data. The worst-case scenario for the modified bisecting algorithm is $2n$ similarity evaluations for $n + 1$ clusters. The best case scenario for the algorithm is $2n$ similarity evaluations for $2^n$ clusters. Of course, there are minor modifications if one wants to do division according to an arbitrary number of clusters at each splitting stage (as opposed to the binary splitting considered here). For example, if we split into $m$ clusters at each node, the worst case scenario is $m n$ similarity evaluations for $(m-1)(n-1) + m$ clusters, and the best case scenario has $m^n$ clusters. If one simply wants to get the most number of clusters for a fixed number of similarity evaluations, this will depend on the details of splitting distribution; however, in the case of optimal partitioning at each splitting level the ideal splitting is $m = 3$ since if we keep the number of clusters fixed at a constant $c$, e.g. $m^n = c$, then minimizing the number of similarity evaluations $M N$ would be the same as minimizing $g(m) \equiv \log(c) m / log(m)$ for integers $m$, in which case the optimal number is $m = 3$, independent of $c$. 

\section{Application: Lorenz Equations} 
\label{sec:lorenz}
We generate a trajectory of the Lorenz equations by starting with the initial condition 
\begin{align*}
(x(0),y(0),z(0)) = (1.4237717232359446, 1.778970017190979, 16.738782836244038)
\end{align*}
and evolving the equations forward in time with a Runge-Kutta 4 time-stepping scheme \cite{butcher2016numerical}, with a timestep size of $\Delta t = 10^{-3}$ for final time $T = 10^5$ leading to $N_t = 10^8$ data points. Thus the data matrix is a long rectangular matrix $X \in \mathbb{R}^{N_s \times N_t}$ where $N_s = 3$. To incorporate symmetries when applying k-means and constructing operators, we take the same time series and multiply the $x$ and $y$ variables by $-1$. 

We then partition state space by setting a threshold probability of $p = 1.4 \times 10^{-6}$ and choosing a binary partition at each stage, yielding roughly $\mathcal{O}(10^6)$ cells at the finest resolution. We then coarse-grain the fine partition according to a probability threshold down to four cells at the coarsest setting. This way, we can have a consistent notion of convergence when refining the partition. Figure \ref{fig:tree} shows a visualization of the resulting hierarchy of partitions. 

The modified bisecting k-means clustering algorithm produces a classifier $\mathcal{C}: \mathbb{R}^3 \rightarrow \mathbb{N}$ that maps a state $\mathbf{s}$ of the Lorenz system to an integer. Upon converting the three-dimensional numbers $\mathbb{R}^3$ into integers, we then use the data-driven construction outlined in \cite{Souza_2024_1, Souza_2024_2} to construct a sparse matrix representation of the infinitesimal generator, $Q$. We also do an analogous calculation for the Perron-Frobenius operator, which may be viewed as the matrix exponential of the generator (although there are subtle differences in their data-driven construction, see \cite{Souza_2024_1, Souza_2024_2}). 

The resulting operators are sparse and we use inverse iteration to compute eigenvalues and eigenvectors 
\cite{golub2013matrix, trefethen1997numerical}. The inverse is calculated iteratively using an incomplete LU factorization. We use the Julia library's default solvers when the matrices are small (less than $2000 \times 2000$). 

The k-means algorithm produces ``cell-centers" (or cluster centers) which, when combined with the zeroth eigenvalue of the generator (eigenvalue one of the Perron-Frobenius operator), are used to calculate statistical properties; see \cite{Souza_2024_1, Souza_2024_2} for details, which will be outlined here. Each cell-center $\sigma^{[n]}$ is associated with a cell $n$,  and a probability $p_n$. The time series is sampled at discrete times $t_n$ from the state $\pmb{s}$. 

A statistic of an observable $g$ is then calculated in two different manners: The first using the data set which corresponds to a temporal average $\langle g \rangle_T$ and the second using the data-driven approximation to the generator, cell centers $\sigma^{[i]}$, and probabilities $p_i$, in symbols $\langle g \rangle_E$:
\begin{align}
\langle g \rangle_T &\equiv \frac{1}{N_t} \sum_{n = 1}^{N_t} g(\pmb{s}(t_n)) 
\\
\label{time_autocorrelation_approximation}
R_T(g, \tau) 
&\equiv  \frac{1}{N_t'} \sum_{n = 1}^{N_t'} g(\pmb{s}(t_n + \text{round}(\tau / \Delta t) \Delta t)) g(\pmb{s}(t_n)) 
\end{align}
where the $\text{round}$ function computes the closest integer and $N_t' = N_t - \text{round}(\tau / \Delta t)$.
The ensemble average equivalents are 
\begin{align}
    \langle g \rangle_E &\equiv \sum_n g( \mathbf{\sigma}^{[n]} ) p_n 
\end{align}
and
\begin{align}
\label{ensemble_autocorrelation_approximation}
    R_E(g, \tau) 
    &\equiv  
    \sum_{n = 1}^{N} g(\mathbf{\sigma}^{[n]}) p_n \left[ \sum_{m=1}^{N}  g( \mathbf{\sigma}^{[m]})  [\exp(Q \tau)]_{mn}  \right].
\end{align}
The operator $[\exp(Q \tau)]_{mn}$ can either be interpreted as the Perron-Frobenius operator at timescale $\tau$, 
\begin{equation}
    \text{pow}\left( \mathcal{P}_{(\Delta \tau)}, \text{round}(\tau / \Delta t) \right) \ , 
\end{equation}
or the matrix exponential of the generator $Q$. See details in \cite{Souza_2024_1, Souza_2024_2} for the assumptions underlying these formulae.

\subsection{Results}
\label{sec:results}

\begin{figure}
\begin{center}
\includegraphics[width=1.0\textwidth]{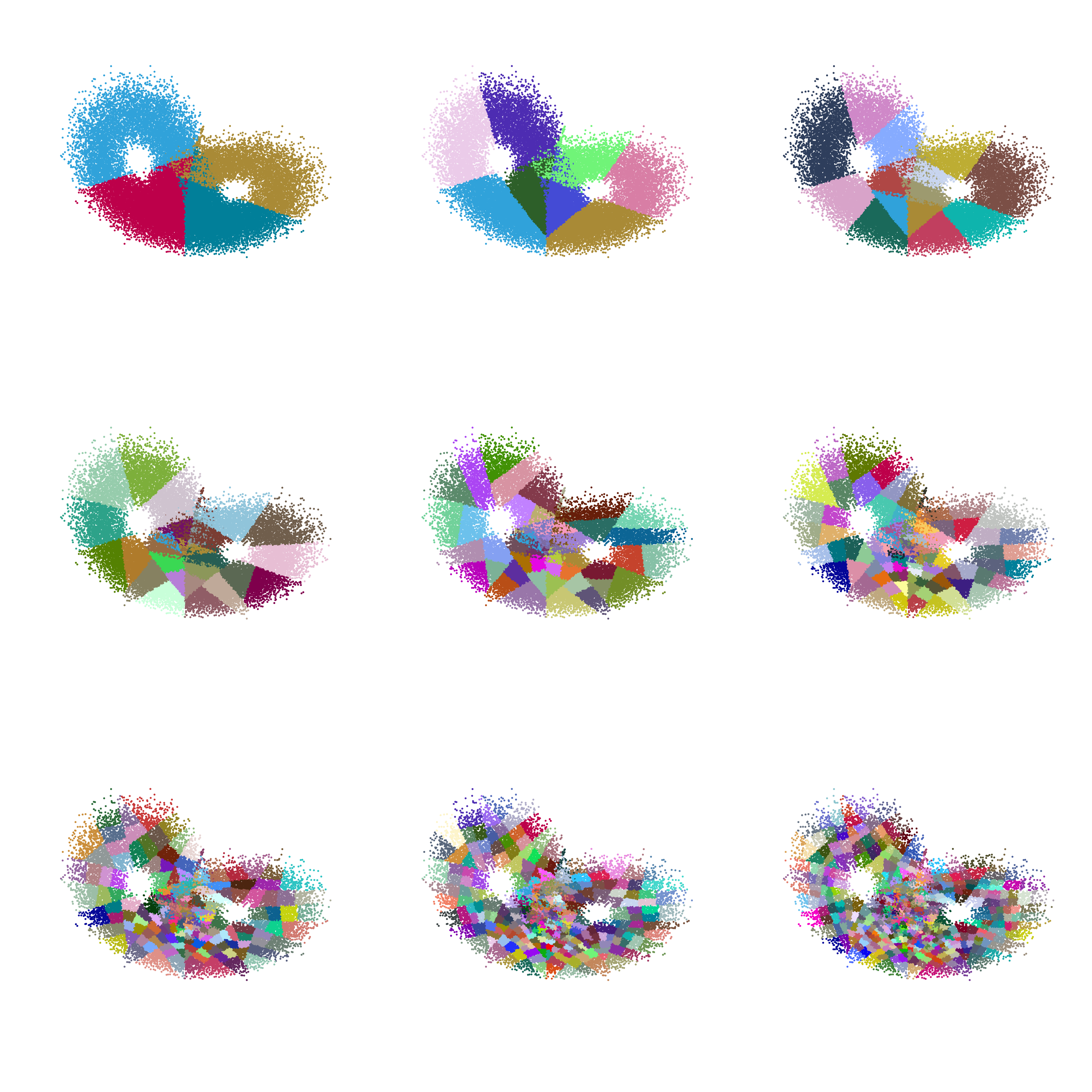}
\caption{\textbf{Hierarchical Modified Bisecting K-Means Algorithm on the Lorenz Attractor.} Here, we show nine different levels of refinement of the Lorenz attractor, starting with four cells (top right) and continually performing bisections based on increasingly small thresholds proceeding leftward and downward. }
\label{fig:butterfly_mosaic}
\end{center}
\end{figure}

Applying the algorithm of Section \ref{sec:algorithm} and methodology of Section \ref{sec:lorenz}, we create a nonlinear dictionary of over 1,000,000 terms. The modified bisecting k-means algorithm yields partitions of the form Figure \ref{fig:butterfly_mosaic}. As we refine from four partitions (top left) to the right-adjacent (top middle) panel, we see a factor of two in the number of cells, increasing from four to eight. A similar increase is found in the subsequent panels; each cell has roughly the same probability. Here, we only show eight levels of refinement from a four-cell partition.

\begin{figure}
\begin{center}
\includegraphics[width=1.0\textwidth]{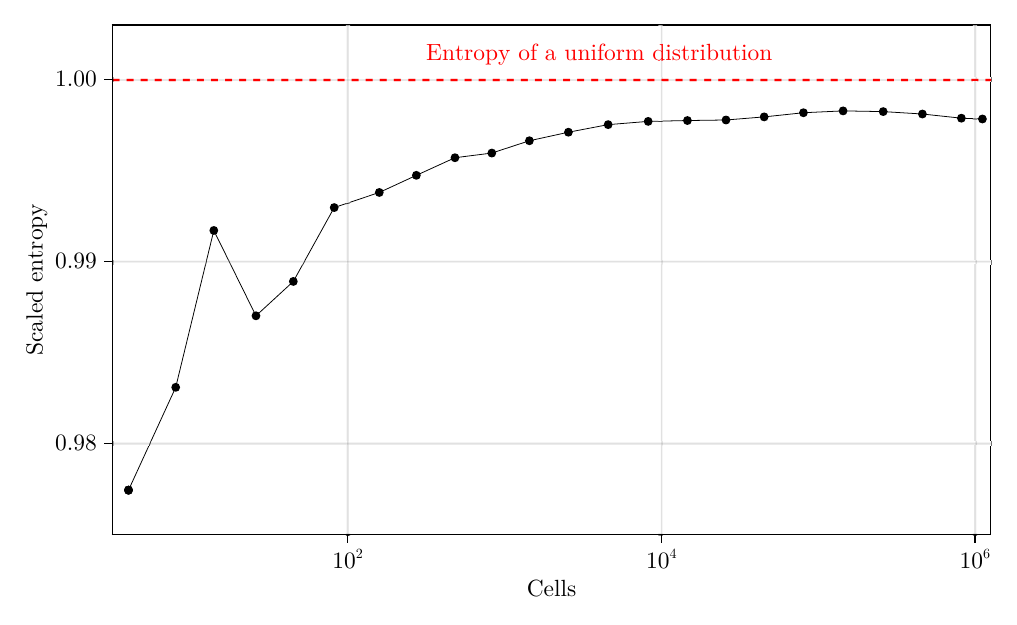}
\caption{\textbf{Scaled Entropy as a Function of Cell Numbers.} The dots represent the log of the scaled entropy, and the red dashed line is the upper bound based on the uniform distribution. The entropy remains close to ideal as the cell numbers increase. }
\label{fig:scaled_entropy}
\end{center}
\end{figure}

We confirm that the scaled entropy, defined as the entropy of the steady-state distribution of the partition (as given by the 0'th eigenvalue of the generator) divided by the uniform distribution's entropy 
\begin{align}
\text{scaled entropy}(p_1, ..., p_N) = \frac{-\sum_{j = 1}^N p_j \log p_j }{-\sum_{j = 1}^N N^{-1} \log N^{-1}}
\end{align}
are similar in Figure \ref{fig:scaled_entropy} for a partition with $N$ cells. The largest value of the scaled entropy is one; thus, the upper bound is zero when taking the logarithm. Being nearly uniform is beneficial to the data-driven method since matrix entries require sufficient samples to estimate exit probabilities and holding times correctly. In other words, regions with much lower probabilities than others will constitute poorly sampled columns of the resulting linear operator. 

\begin{figure}
\begin{center}
\includegraphics[width=1.0\textwidth]{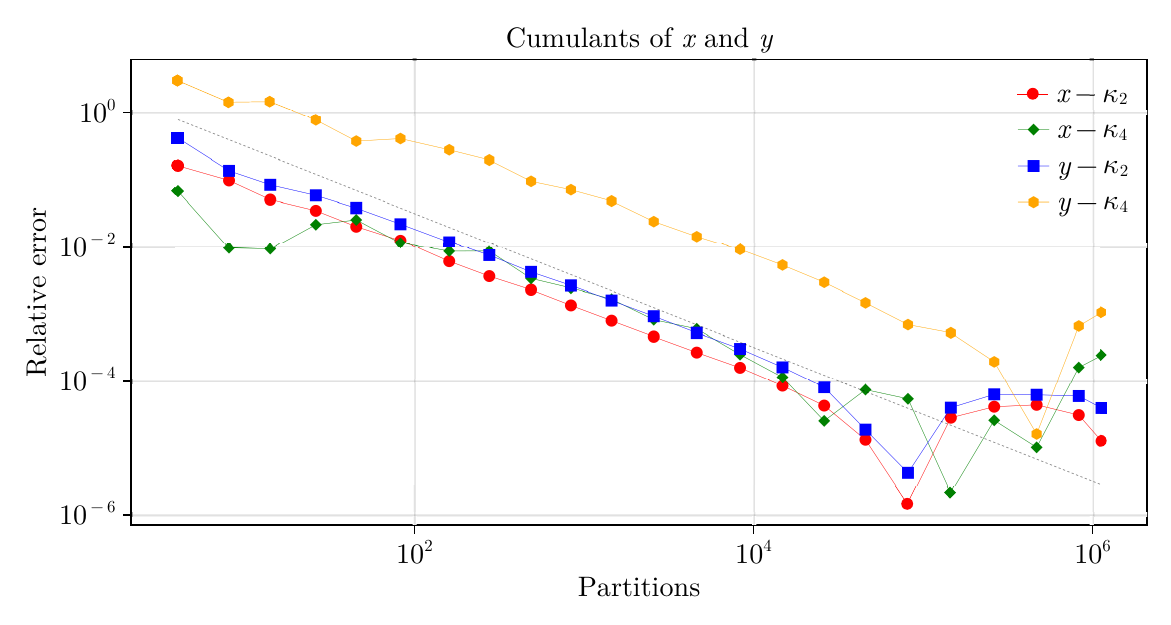}
\caption{\textbf{Convergence of Cumulants of the $x$ and $y$ Variables.}. Here, we compare the relative error of the statistical model with a temporal average of the data as we increase the number of cells in the statistical model. We see saturated convergence around $10^5$ cells. }
\label{fig:cumulant_convergence_xy}
\end{center}
\end{figure}

\begin{figure}
\begin{center}
\includegraphics[width=1.0\textwidth]{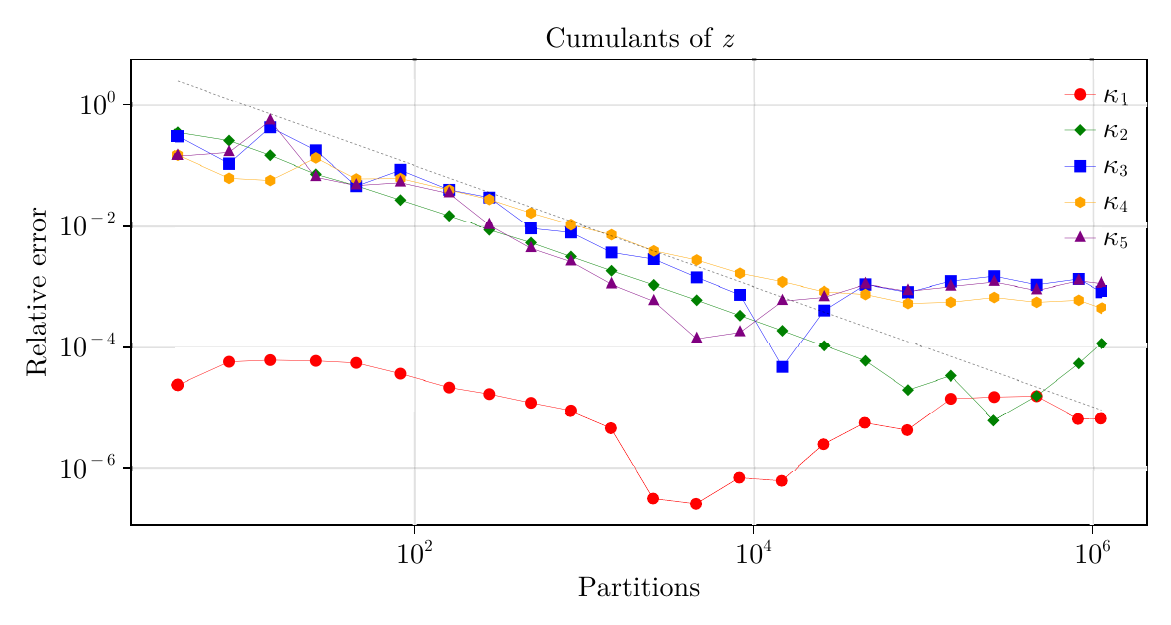}
\caption{\textbf{Convergence of Cumulants of the $z$ Variable.}. Here, we compare the relative error of the statistical model with a temporal average of the data as we increase the number of cells in the statistical model. We see saturated convergence around $10^5$ cells. }
\label{fig:cumulant_convergence_z}
\end{center}
\end{figure}

\begin{figure}
\begin{center}
\includegraphics[width=1.0\textwidth]{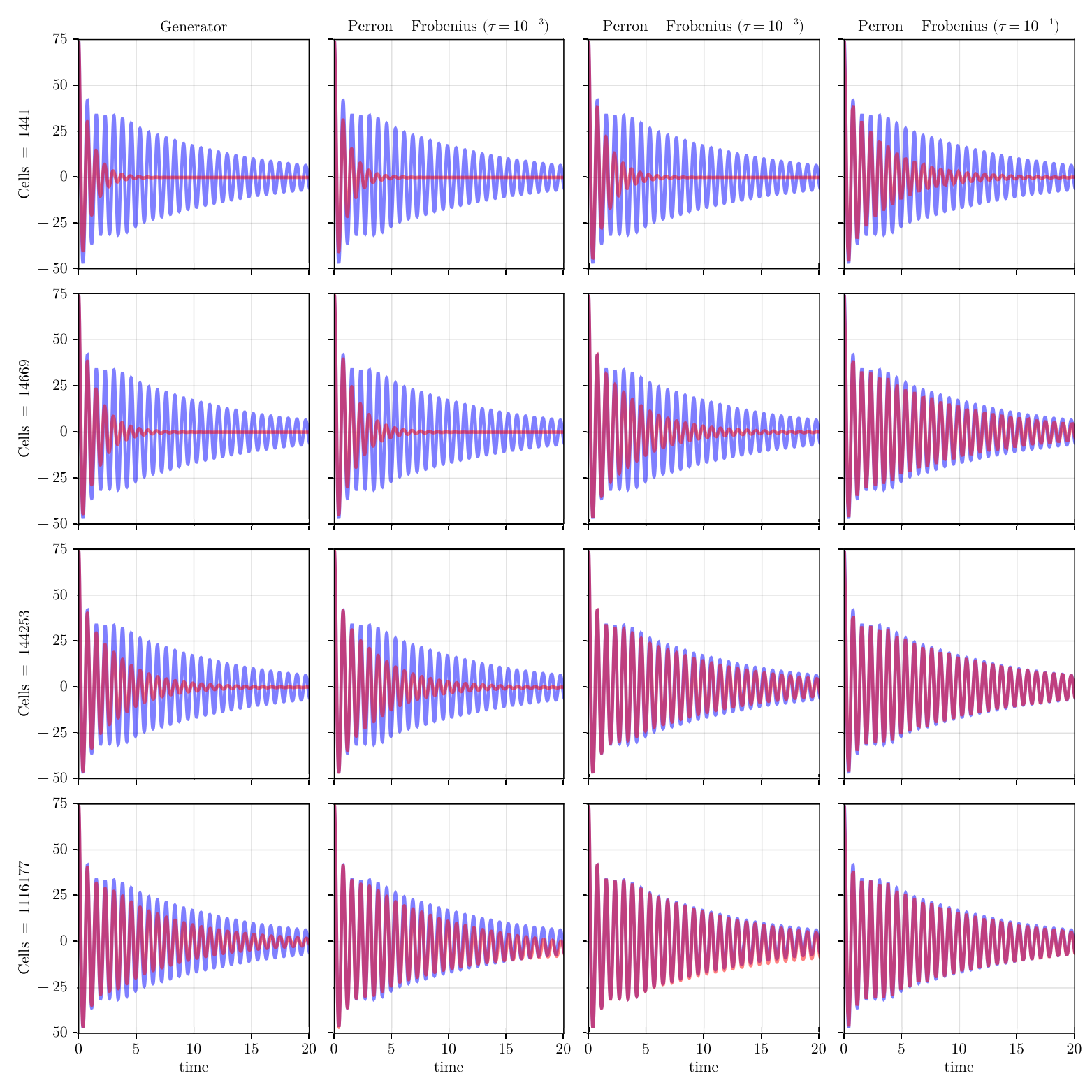}
\caption{\textbf{Autocorrelation of the $z$ Variable as a Function of Cell Number and Data-Driven Method}. Here, we show that as we increase the number of cells, we converge to the $z$-autocorelation determined from the time series.}
\label{fig:zautocorrelation}
\end{center}
\end{figure}

We further check the ability of the data-driven method to replicate steady-state statistics of the non-zero cumulants in Figures \ref{fig:cumulant_convergence_xy} and \ref{fig:cumulant_convergence_z}. In both cases, the algorithm begins to saturate in accuracy for steady-state values at around $10^5$ cells, perhaps due to insufficient sampling for estimating the temporal average. (We will show later that the other eigenvalues appear to increase in accuracy even for $10^6$ cells.) Indeed, we evolved the Lorenz equations until $T = 10^5$. From our estimates of the longest decorrelation time of the generator (see Figure \ref{fig:zautocorrelation}), we have between $10^3$ and $10^4$ decorrelated in time samples over this period. The empirical error in estimating the cumulants should begin dominating the inaccuracy at around three digits. Before reaching this saturation, the data-driven method appears to exhibit a first-order convergence rate concerning refining the number of cells, as exhibited by the dashed line in both cases. This convergence rate is valid for all the cumulants except for the first cumulant (the average value) of the $z$-variable, where the data-driven method accurately estimates the value, even for a small number of cells.

To investigate the convergence of the other eigenvalues of the Lorenz attractor is more challenging. Firstly, it is unclear whether eigenvalues and eigenvectors of the Lorenz attractor are meaningful or well-defined in the limit of the ever-increasing resolution of the deterministic Lorenz equations. The data-driven method introduces a stochastic regularization in the estimation of the operator \cite{Souza_2024_1, Souza_2024_2}; however, specific quantities are well-posed and can be computed in different manners that serve as a comparison in the present case. In particular, the autocorrelation of any observable of a system can be decomposed according to the eigenvectors and eigenvalues of the Perron-Frobenius operator or generator via the formulas in \cite{Souza_2024_1, Souza_2024_2} and repeated in Section \ref{sec:lorenz}. Thus, we show the calculation with an ever-increasing number of cells (red) concerning the $z$-autocorrelation computed using temporal averages (blue) in Figure \ref{fig:zautocorrelation}, where the purple denotes an overlap between the two different methods of calculation. In each column, we construct the transfer operator in four ways: the infinitesimal generator and the Perron-Frobenius operator with three different timescales. Each row corresponds to an order of magnitude increase in the number of cells for a given partition, starting from $\mathcal{O}(10^3)$ cells and ending with $\mathcal{O}(10^6)$ cells. We see that at lower resolutions, the damping due to the loss of information of state-space initial conditions is quite substantial but ameliorated as one examines operators over longer timescales. Despite the over-damping, one also sees that the method captures the oscillatory timescale appropriately. This observation may be related to the example in Appendix D of \cite{Souza_2024_1, Souza_2024_2}, where the imaginary eigenvalues converged at a second-order rate. As resolution increases, each of the separate methods of calculating the autocorrelation becomes closer to the ``ground-truth" temporal autocorrelation, with the Perron-Frobenius operator estimated to be the longest timescale converging at the fastest rate. This convergence shows that all the operators' eigenvalues conspire appropriately to produce the correct autocorrelation for the $z$ observable. Next, we look at what happens to an individual eigenfunction of the operator.

\begin{figure}
\begin{center}
\includegraphics[width=1.0\textwidth]{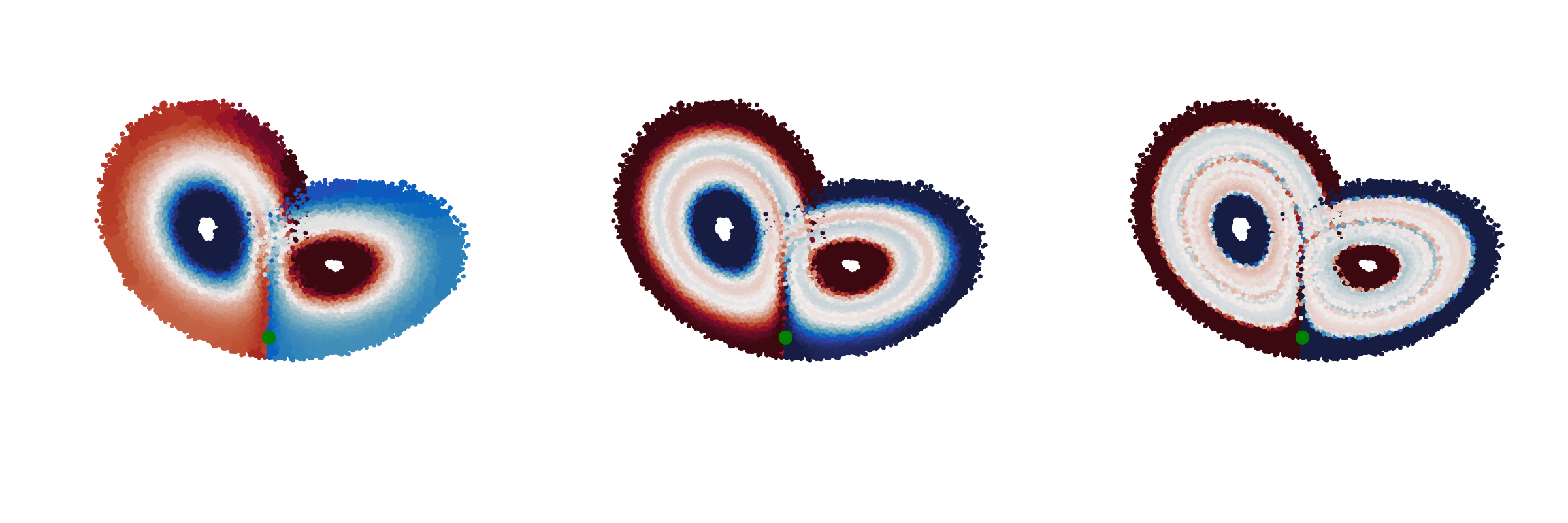}
\caption{\textbf{Quasi-Invariant Set Koopman Eigenfunction State Space.} Here, we show Koopman eigenfunctions associated with the Quasi-Invariant set of the Lorenz equations (second Largest real eigenvalue), numerically continued from $\mathcal{O}(10^4) \rightarrow  \mathcal{O}(10^5) \rightarrow \mathcal{O}(10^6)$ cells, respectively, from left to right. We see finer detail structures emerge as resolution increases. The eigenfunctions here are constructed from the generator's left eigenvectors. The green circle highlights the trajectory at time $t = 5$.}
\label{fig:koopman_state}
\end{center}
\end{figure}

\begin{figure}
\begin{center}
\includegraphics[width=1.0\textwidth]{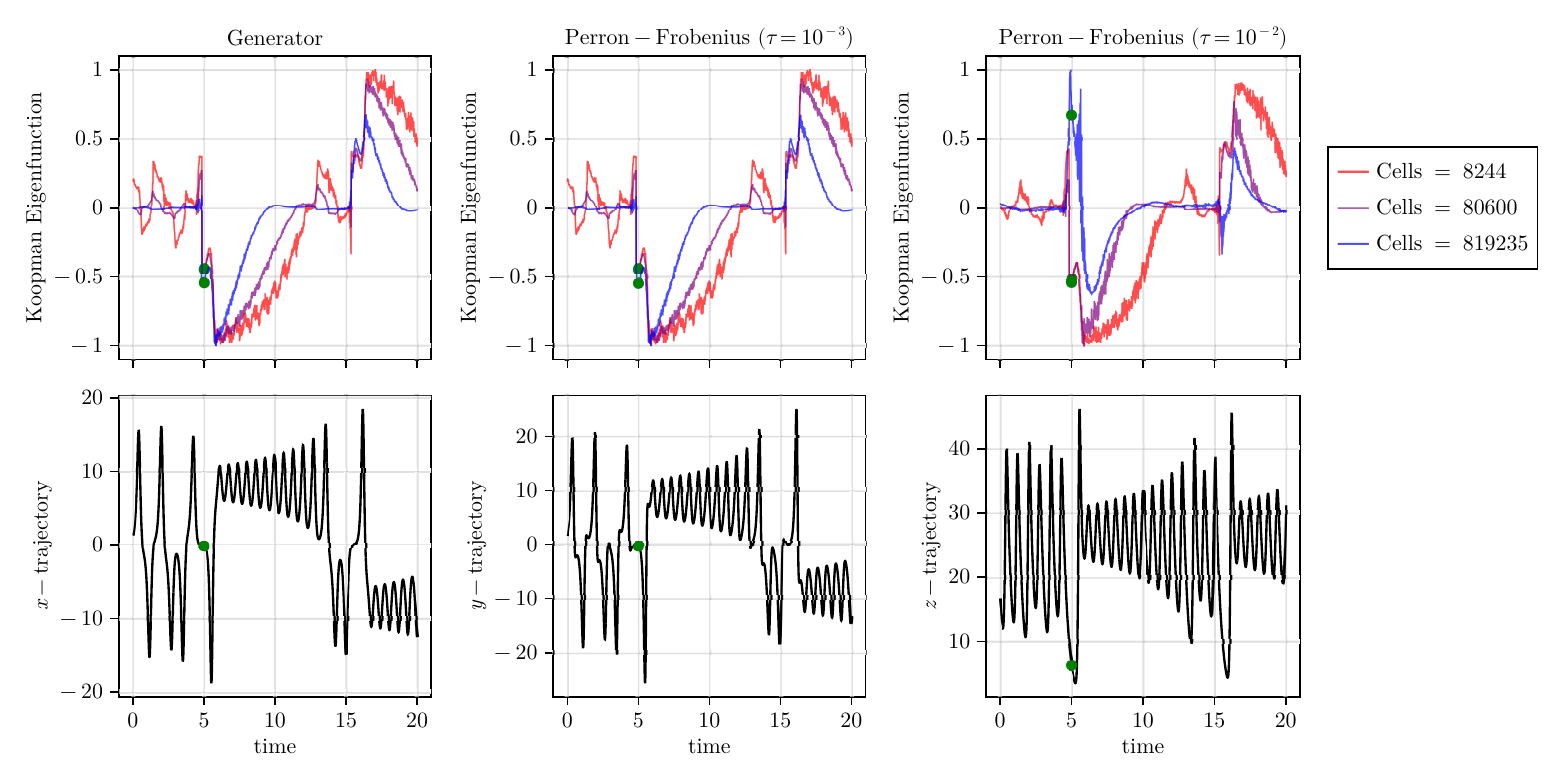}
\caption{\textbf{Quasi-Invariant Set Koopman Eigenfunction Timeseries.} Similar as Figure \ref{fig:koopman_state} but as a function of time. We see that the Koopman eigenfunctions are irregular as a function of time and become increasingly intermittent as resolution increases. Furthermore, we see correspondence between the different ways of constructing operators, with the Perron-Frobenius for $\tau = 10^{-2}$ providing an increased ``effective" resolution for the Koopman eigenfunction. The green dots show the trajectory and the value of the Koopman eigenfunction at $t = 5$.}
\label{fig:koopman_time}
\end{center}
\end{figure}

As mentioned in the previous paragraph, the autocorrelations use the other eigenvalues shown in Figures \ref{fig:koopman_state} and \ref{fig:koopman_time}. We isolate the Koopman eigenfunction associated with the quasi-invariant set of the Lorenz equations. This choice was determined by computing the second-largest pure real eigenvalue of the data-constructed infinitesimal generator (the largest real eigenvalue is zero) and computing the left-eigenvector associated with this eigenvalue via inverse iteration \cite{golub2013matrix, trefethen1997numerical}. We comment that there are imaginary eigenvalues whose real part is larger and thus decorates more slowly. We visualize this eigenfunction in state space in Figure \ref{fig:koopman_state}. The blue and red regions serve as the main divisors and are locations where a trajectory of the Lorenz equations stays for extended periods. These regions essentially correspond to trajectors originating near the z-axis. These trajectories then sweep down close to the origin before making a large excursion near the exterior of the Lorenz attractor and transferring to the opposite lobe. The underlying guide for the dynamics in this case can be isolated to the heteroclinic connections between the fixed points of the Lorenz equations. As we increase resolution, finer-scale structures emerge in the interior lobes of the Lorenz attractor. In other words, the Koopman eigenfunction becomes increasingly intermittent as we increase resolution. At the lowest resolution (the left-most plot in Figure \ref{fig:koopman_state}), the white regions essentially correspond to the location of the fundamental periodic orbit of the Lorenz equations.

In Figure \ref{fig:koopman_time}, we visualize the same Koopman eigenfunction from Figure \ref{fig:koopman_state} as a function of time by associating the dynamics (bottom row) at each point in state space with the corresponding value of the eigenfunction (top row). We do this for three different estimates of the operator (the columns) at three different resolutions ($\mathcal{O}(10^4)$ cells in red, $\mathcal{O}(10^5)$ cells in purple, $\mathcal{O}(10^6)$ cells in blue). At around $t \approx 5$, we see a spike in all estimates of the Koopman eigenfunctions. This spike is an early warning indicator that a transition to the quasi-invariant set is occurring. This spike in the Koopman eigenfunction can be interpreted dynamically by examining the time series of the Lorenz equations in the bottom row. After $t \approx 5$, one sees a long negative $x$ and $y$ excursion (as well as small values of $z$) followed by a long oscillatory period in the $x$ and $y$ variables where transitions to different lobes do not occur. Similarly, the spike in the opposite direction at time $t \approx 15 $ indicates a similar feature but a transition from the positive to negative lobe. As a function of resolution, one sees that the eigenfunction becomes increasingly intermittent as resolution increases. Furthermore, one can see a higher effective resolution for this eigenfunction in the case of the operator constructed over a larger timescale. In particular one sees that the $\mathcal{O}(10^5)$ cell case of the $\tau = 10^{-2}$ operator is similar to that of the $\mathcal{O}(10^6)$ cell case of the $\tau = 10^{-3}$ perron-frobenius  operator and generator. Lastly, we comment that the spike in the Koopman eigenfunction tends to coincide with regions in state space where $0 \approx \dot{x} \approx \dot{y} \approx x \approx y \approx z $. 

\section{Conclusions}
\label{sec:conclusions}

In this study, we discretized the Perron-Frobenius operator of the Lorenz equations using a modified bisecting k-means algorithm. We investigated the effect of increasing the number of clusters (i.e., the number of terms in the nonlinear dictionary of the extended dynamic mode decomposition) up to 1,000,000 clusters with respect to the ability to represent steady-state statistics of the state variables, autocorrelations, and a Koopman eigenfunction. We partially addressed some of the conceptual questions from the introduction: 
\begin{enumerate}
    \item What does it mean to approximate the Koopman eigenfunctions of a chaotic dynamical system in the absence of noise? \textit{It seems that increasing the number of terms in the dictionary yields better representations of steady-state statistics, dynamics, and temporal autocorrelations; however, the eigenfunctions themselves seem to become increasingly fractal and it is difficult to assess in what sense the eingefunctions themselves are ``converging".} 
    \item To what extent does increasing the number of terms of a nonlinear dictionary ``help"? \textit{If one is after statistical information, there is no added information beyond the data using a piecewise constant basis in the nonlinear dictionary. There is a compromise on the amount of data available and the ability to estimate the operator's entries.}
    \item At what timescale should a Koopman operator be constructed? \textit{It seems like a naive construction of the infinitesimal generator is more dissipative than the Perron-Frobenius operator at larger timescales; however, Perron-Frobenius operators constructed at timescales much larger than the Koopman eigenvalue of interest quickly lose meaning. Thus the best strategy is should be iterative in the absence of system information. That is, start with the generator to obtain an estimate of the timescales involved, then directly construct a Perron-Frobenius operator that is constructed on timescales that don't ``skip over" the target timescale. }
\end{enumerate}

Extending the work here to more complex and higher-dimensional dynamical systems is an interesting future direction. As shown in \cite{Souza_2024_1, Souza_2024_2}, it is possible to construct a data-driven representation of the statistics, even in the high-dimensional setting; however, to effectively apply the methodology listed herein, it is necessary first to perform a data-reduction technique such as SVD/PCA/EOF \cite{lorenz1956empirical} or autoencoders \cite{Goodfellow-et-al-2016}. In addition, one can use other data-driven constructions of the infinitesimal generator, such as that of \cite{giorgini2023clustering}, to better approximate generators over many timescales. Extensions that can use a similar hierarchical approach to clustering could use a physics-based partitioning (e.g., \cite{Jiménez_2023}), then further subdivide using additional physics criteria or the k-means algorithm presenter here.

The goals of applying the extended dynamic mode decomposition are manifold. The data-driven method provides a way of computing statistical eigenfunctions, which often serve as a method for obtaining ``early warning indicators" for interesting features in a dynamical system. Additionally, they serve as the necessary ingredients in certain theoretical relations, such as the calculation of turbulent transport operators \cite{Souza_Lutz_Flierl_2023} or response functions \cite{markov_response_functions}. The methodology outlined here provides an additional tool for addressing challenges in complex systems research.

\appendix

\section*{Acknowledgments}
This work acknowledges support by Schmidt Sciences, LLC, through the Bringing Computation to the Climate Challenge, an MIT Climate Grand Challenge Project. We thank Peter J. Schmid for introducing us to the algorithms and data structures for constructing unstructured trees. 

\bibliographystyle{siamplain}
\bibliography{references}
\end{document}